\documentclass[sigplan,screen,noacm=true]{acmart}
\usepackage{balance}
\usepackage{caption,subcaption}
\usepackage{float}
\usepackage{textcomp}
\usepackage{multicol}
\usepackage{enumitem}

\usepackage{subfiles}
\usepackage[ruled,vlined]{algorithm2e}

\SetCommentSty{mycommfont}

\usepackage{color,xcolor}
\usepackage{listings}
\usepackage{tikz,pgfplots}
\pgfplotsset{compat=newest}
\usetikzlibrary{
  automata,
  arrows,
  arrows,decorations,
  decorations.pathreplacing,
  backgrounds,
  calc,
  fit,
  patterns,
  positioning,
  petri,
  shapes,
  shapes.multipart,
  shapes.symbols,
  shadows.blur,
  tikzmark
}

\tikzset{
  add path/.style = {
    decoration={show path construction,
      moveto code={
        \xdef\savedpath{\savedpath (\tikzinputsegmentfirst)}
      },
      lineto code={
        \xdef\savedpath{\savedpath -- (\tikzinputsegmentlast)}
      },
      curveto code={
        \xdef\savedpath{\savedpath .. controls (\tikzinputsegmentsupporta)
          and (\tikzinputsegmentsupportb) ..(\tikzinputsegmentlast)}
      },
      closepath code={
        \xdef\savedpath{\savedpath -- cycle}
      }
    },
    decorate
  },
  store path/.style = {add path},
  store path/.prefix code={\xdef\savedpath{}},
  callouts/.style={
    store path,
    append after command={
       foreach \target in {#1}{
        ($(callout)!2pt!-90:\target$)--\target --($(callout)!2pt!90:\target$)
      } \savedpath
    },
    alias=callout
  },
  custom style/.style={fill=blue!20,text=,blur shadow},
  cloudy/.style={cloud,cloud puffs=10,cloud puff arc=120, aspect=2}
}

\makeatletter
\pgfkeys{/tikz/.cd,
  dogear size/.initial=5pt,
}
\pgfdeclareshape{dogeared}{ 
  \inheritsavedanchors[from=rectangle]
  \inheritanchorborder[from=rectangle]
  \inheritanchor[from=rectangle]{center}
  \inheritanchor[from=rectangle]{north}
  \inheritanchor[from=rectangle]{south}
  \inheritanchor[from=rectangle]{west}
  \inheritanchor[from=rectangle]{east}

  \savedmacro\dogearsize{%
    \edef\dogearsize{\pgfkeysvalueof{/tikz/dogear size}}%
  }

  \backgroundpath{
    \southwest \pgf@xa=\pgf@x \pgf@ya=\pgf@y
    \northeast \pgf@xb=\pgf@x \pgf@yb=\pgf@y
    \pgf@xc=\pgf@xb \advance\pgf@xc by-\dogearsize
    \pgf@yc=\pgf@yb \advance\pgf@yc by-\dogearsize
    \pgfpathmoveto{\pgfpoint{\pgf@xa}{\pgf@ya}}
    \pgfpathlineto{\pgfpoint{\pgf@xa}{\pgf@yb}}
    \pgfpathlineto{\pgfpoint{\pgf@xc}{\pgf@yb}}
    \pgfpathlineto{\pgfpoint{\pgf@xb}{\pgf@yc}}
    \pgfpathlineto{\pgfpoint{\pgf@xb}{\pgf@ya}}
    \pgfpathclose
  }
  \foregroundpath{
    \southwest \pgf@xa=\pgf@x \pgf@ya=\pgf@y
    \northeast \pgf@xb=\pgf@x \pgf@yb=\pgf@y
    \pgf@xc=\pgf@xb \advance\pgf@xc by-\dogearsize
    \pgf@yc=\pgf@yb \advance\pgf@yc by-\dogearsize
    \pgfpathmoveto{\pgfpoint{\pgf@xc}{\pgf@yb}}
    \pgfpathlineto{\pgfpoint{\pgf@xc}{\pgf@yc}}
    \pgfpathlineto{\pgfpoint{\pgf@xb}{\pgf@yc}}
    \pgfpathclose
  }
}
\makeatother

\newcommand*\circled[1]{\tikz[baseline=(char.base)]{
    \node[shape=circle,draw,inner sep=1pt] (char) {#1};}}

\newcommand{\jit}{JIT}
\newcommand{\vm}{VM}

\newcommand{\rpython}{RPython}
\newcommand{\truffle}{Truffle/Graal}

\newcommand{\hotspot}{HotSpot\textsuperscript{\texttrademark}}
\newcommand{\lif}{language implementation framework}
\newcommand{\Lif}{Language implementation framework}

\newcommand{\memo}[1]{}
\newcommand{\dotracestitching}{\textsc{DoTraceStitching}}
\newcommand{\tla}{\textsc{tla}}

\setcopyright{none}
\copyrightyear{2022}
\acmYear{2022}
\acmDOI{}

\acmConference[PEPM '22]{PEPM '22: ACM SIGPLAN Workshop on Partial Evaluation and Program
  Manipulation}{January 17--18, 2022}{Philadelphia, Pennsylvania, United States}
\acmBooktitle{PEPM '22: ACM SIGPLAN Workshop on Partial Evaluation and Program
  Manipulation, January 17--18, 2022, Philadelphia, Pennsylvania, United States}
\acmPrice{}
\acmISBN{}



\title{Two-level Just-in-Time Compilation with One Interpreter and One Engine}
\author{Yusuke Izawa}
\email{izawa@prg.is.titech.ac.jp}
\affiliation{
    \institution{Tokyo Institute of Technology}
    \state{Tokyo}
    \country{Japan}
}

\author{Hidehiko Masuhara}
\email{masuhara@acm.org}
\affiliation{
    \institution{Tokyo Institute of Technology}
    \state{Tokyo}
    \country{Japan}
}

\author{Carl Friedrich Bolz-Tereick}
\email{cfbolz@gmx.de}
\affiliation{
  \institution{Heinrich-Heine-Universit\"{a}t D\"{u}sseldorf}
  \state{North Rhine-Westphalia}
  \country{Germany}
}

\begin{abstract}
  Modern, powerful virtual machines such as those running Java or JavaScript support
multi-tier JIT compilation and optimization features to achieve their high
performance. However, implementing and maintaining several compilers/optimizers that
interact with each other requires hard-working VM developers. In this paper, we propose a
technique to realize two-level JIT compilation in RPython without implementing several
interpreters or compilers from scratch. As a preliminary realization, we created adaptive
RPython, which performs both baseline JIT compilation based on threaded code and tracing
JIT compilation. We also implemented a small programming language with it. Furthermore,
we preliminarily evaluated the performance of that small language, and our baseline JIT
compilation ran 1.77x faster than the interpreter-only execution. Furthermore, we
observed that when we apply an optimal JIT compilation for different target methods, the
performance was mostly the same as the one optimizing JIT compilation strategy, saving
about 40 \% of the compilation code size.

\end{abstract}

\renewcommand{\shortauthors}{Izawa, Masuhara, and Bolz-Tereick.}

\begin{document}

\sloppy

\maketitle

\section{Introduction}
\label{sec:introduction}

\Lif{}s, e.g. \rpython{}~\cite{BOLZ2015408} and
\truffle{}~\cite{Wurthinger:2017:PPE:3062341.3062381}, are tools used to build a virtual
machine (\vm{}) with a highly efficient just-in-time (\jit{}) compiler which is done by
providing an interpreter of the language. For example, PyPy~\cite{Bolz2009}, which is a fully
compatible Python implementation, achieved 4.5x speedup from the original CPython 3.7
interpreter~\cite{pypyspeedcenter}. The other successful examples include
Topaz~\cite{topaz}, Hippy \vm{}~\cite{hippyvm}, TruffleRuby~\cite{truffleruby}, and
GraalPython~\cite{graalpython}.

One of the limitations of \rpython{} and \truffle{}, which is differed from the current
language-specific \vm{}s such as Java or JavaScript, is that they don't support a
multitier \jit{} compilation strategy.
One na\"ive approach for multitier compilation is to create compilers for each
optimization level. However, this approach makes existing implementations more complex and
requires more developing efforts to implement and maintain --- we need to at least
consider how to share components, along with how to exchange profiling information and compiled
code between each compiler. To avoid this situation, we have to find another efficient
and reasonable way to support multitier \jit{} compilation in a framework. Several works
in \rpython{}~\cite{Bolz:2011:RFM:2069172.2069181,Bauman:2015:PTJ:2784731.2784740,Huang2016}
found that specifying special annotations in an interpreter definition can influence how
\rpython{}'s meta-tracing \jit{} compiler works. In other words, we can view the
interpreter definition as a specification of a compiler. We believe those approaches can
be extended to achieve an adaptive optimization system at the meta-level.

As a proof of concept of language-agnostic multitier \jit{} compilation, we propose
\emph{adaptive \rpython{}}. Adaptive \rpython{} can generate a \vm{} with two-level \jit{}
compilation. We do not create two separated compilers in adaptive \rpython{} but indeed generate
two different interpreters --- the one for the first level of compilation and another
is for the second level. As first-level compilation, we support threaded
code~\cite{Bell:1973:10.1145/362248.362270,Hong:10.1145/146559.146561} in a meta-tracing
compiler (we call this technique \emph{threaded code
  generation}~\cite{Izawa:2021:ThreadedCodeGen:unpub}). The second one is \rpython{}'s
original tracing \jit{} compilation. We can switch those compilation tiers by moving between
the two interpreters. In addition, adaptive \rpython{} generates two interpreters
from one definition. It will reduce the implementation costs that language developers
should pay if they are in a traditional development way. In the future, we plan to
extend this system to realize a nonstop compilation-tier changing mechanism in a \lif{}
--- we can use an appropriate and efficient optimization level or a compilation strategy,
depending on the executed program.

The contributions of the current paper can be summarized as follows:

\begin{itemize}
\item An approach to generate multiple interpreter implementations from one common
  interpreter definition to obtain \jit{} compilers with different optimization levels.
\item The technical details of enabling threaded code generation as a first-level \jit{}
  compilation by driving an existing meta-tracing \jit{} compiler engine.
\item The preliminary evaluation of our two-level \jit{} compilation on a simple
  programming language.
\end{itemize}

The rest of current paper is organized as follows:
Section~\ref{sec:two_level_jit_with_rpython} proposes an idea and technique to support
two-level \jit{} compilation with one interpreter in \rpython{} without crafting a
compiler from scratch. We evaluate our preliminary implementation in
Section~\ref{sec:preliminary_eval}. Section~\ref{sec:related_work} discusses related work,
and we conclude the paper and denote the future work in
Section~\ref{sec:conclusion_future_work}.


\section{Two-level Just-in-Time Compilation with RPython}
\label{sec:two_level_jit_with_rpython}

\begin{figure}[t]
  \centering
  \begin{tikzpicture}[auto,transform shape,scale=0.75,remember picture,inner sep=2pt,align=center]
    \tikzstyle{block} = [draw,rectangle,text width=20mm,]
    \tikzstyle{native} = [draw,rounded rectangle,text width=10mm,minimum height=7.5mm,fill=gray!10]
    \tikzstyle{markplace} = [rectangle callout,]
    \tikzstyle{data flow} = [->,thick]
    \tikzstyle{exec flow} = [->,dotted,very thick]

    \node at (0,0) [block,dogeared,text width=35mm,minimum height=7.5mm,inner sep=2pt] (generic) {\bf Generic Interpreter};
    \node [block,below=5mm of generic,text width=30mm,minimum height=7.5mm,fill=blue!20,inner sep=2pt] (adaptiverpy) {\bf Adaptive RPython};

    \node at (-5,-4.5) [block,dogeared,text width=15mm,] (base prog)
    {base-program};

    \node [block,right=10mm of base prog] (baseline interp) {baseline JIT \\ interp.};
    \node [block,right=10mm of baseline interp] (tracing interp) {tracing JIT \\
      interp.};

    \node [native,below=7.5mm of baseline interp,] (baseline native) {native};
    \node [native,below=7.5mm of tracing interp,] (tracing native) {native};

    \draw [data flow]  (base prog) -- (baseline interp);

    \draw [exec flow] (baseline interp) to [loop above] node [auto] {} (1);

    \draw [data flow] (baseline interp) -- (baseline native);

    \draw [exec flow] (baseline native) to [loop right] node [auto] {} ();

    \path [data flow] (baseline interp.east) edge [->,bend left=40,midway] node (switch) {}
    (tracing interp.west);

    \draw [exec flow] (tracing interp) to [loop above] node [auto] {} ();

    \draw [data flow] (tracing interp) -- (tracing native);

    \draw [exec flow] (tracing native) to [loop right] node [auto] {} ();

    \node [draw,rectangle callout,callout absolute pointer=(switch),below=25mm of
    switch,fill=orange!50] (callout) {tracing-JIT-suitable \\ hot spot found};

    \node [draw,fill=white,rectangle callout,callout absolute pointer=(baseline
    native.west),left=5mm of baseline native,font=\small,] {method-\\based};

    \draw [very thick] (generic) -- (adaptiverpy);
    \draw [->,very thick] (adaptiverpy.south) -- (baseline interp.45);
    \draw [->,very thick] (adaptiverpy.south) -- (tracing interp);

    \draw [dotted,very thick] ($(current bounding box.west) + (-1.25,1.25)$)
    to node[pos=0.1,above,label=above:{\small VM generation time},label=below:{\small
      run-time}] {} ($(current bounding box.east) + (-.5,1.25)$) {};

    \matrix [below left,anchor=north west,xshift=0mm,yshift=7.5mm,
    cells={nodes={font=\footnotesize,anchor=west}},inner sep=2pt]
    at (current bounding box.south west) {
      \draw [data flow] (0,0) -- ++ (0.6,0); & \node{data flow};\\
      \draw [exec flow] (0,0) -- ++ (0.6,0); & \node{exec. flow};\\
    };
  \end{tikzpicture}
  \caption{An overview of adaptive \rpython{}: two different interpreters are generated
    from a single generic interpreter by adaptive RPython at \vm{} generation time. At
    run-time, the two interpreters that is, the baseline \jit{} and tracing \jit{} interpreters,
    behave as tier 1 and tier 2, respectively.}
  \label{fig:aot_runtime_overview_adaptive_rpy}
\end{figure}
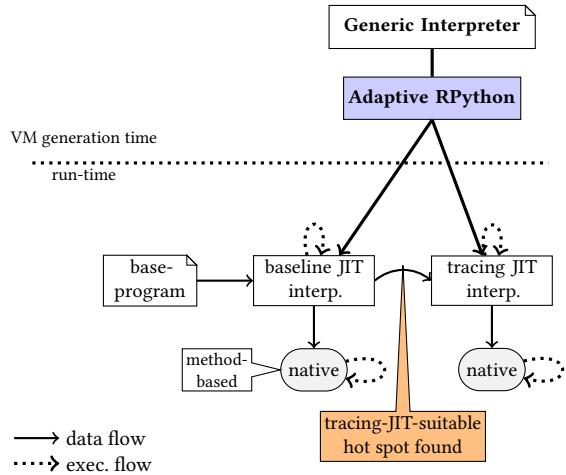

In this section, we propose a multitier meta-tracing \jit{} compiler framework
called \emph{adaptive \rpython} along with its technique to support two different
compilation levels with one interpreter and one engine. Section~\ref{sec:adaptive_rpython}
introduces adaptive \rpython{}. Then, Sections~\ref{sec:generic_interp}
and~\ref{sec:jit_trace_stitching_in_action} explain the technical details to realize
two-level \jit{} compilation in \rpython{}. Sections~\ref{sec:generic_interp}
and~\ref{sec:jit_trace_stitching_in_action} correspond to ``one interpreter'' and ``one
engine,'' respectively.

\subsection{Adaptive RPython}
\label{sec:adaptive_rpython}

Adaptive \rpython{} lets the existing meta-tracing engine to behave in two
ways: as both a baseline \jit{} compiler and a tracing \jit{} compiler. To realize this
behavior, the most obvious approach would be to implement two different compilers or
interpreters. However, this approach increases the amount of implementation necessary.
Thus, we do not craft compilers from scratch but derive two different compilation strategies
by providing a specializing interpreter called the \emph{generic interpreter} to Adaptive
RPython. Figure~\ref{fig:aot_runtime_overview_adaptive_rpy} shows an overview of
how Adaptive \rpython{} and the generic interpreter work. At \vm{} generation time (the
upper half of Figure~\ref{fig:aot_runtime_overview_adaptive_rpy}), the developer writes the
\emph{generic interpreter}. Adaptive \rpython{} generates two different interpreters that
support different \jit{} compilation tiers, e.g. baseline \jit{} and tracing \jit{}
interpreters work as tier-1 and tier-2, respectively. At run-time (the bottom half of
Figure~\ref{fig:aot_runtime_overview_adaptive_rpy}), a generated baseline \jit{} interpreter
firstly accepts and runs a base-program. While running the program, the execution switches to
a generated tracing \jit{} interpreter if the hot spot is suitable for tracing \jit{}
compilation.


\begin{lstlisting}[float=t,language=python,caption={A skeleton of the generic interpreter
    definition.}, label={lst:generic_interp_def}]
jittierdriver = JitTierDriver(pc='pc')

class Frame:
    def interp(self);
        pc = 0
        bytecode = self.bytecode
        jitdriver.jit_merge_point(pc=pc,
            bytecode=bytecode,self=self)
        opcode = ord(bytecode[pc])
        pc += 1
        if opcode == JUMP_IF:
            target = bytecode[pc]
            jittierdriver.can_enter_tier1_branch(
                true_path=target,false_path=pc+1,
                cond=self.is_true)
            if we_are_in_tier2(kind='branch'):
                # do stuff

        elif opcode == JUMP:
             target = bytecode[pc]
             jittierdriver.can_enter_tier1_jump(target=target)
             if we_are_in_tier2(kind='jump'):
                 # do stuff

        elif opcode == RET:
            w_x = self.pop()
            jittierdriver.can_enter_tier1_jump(ret_value=w_x)
            if we_are_in_tier2(kind='ret'):
                # do stuff
\end{lstlisting}

\subsection{Generic Interpreter}
\label{sec:generic_interp}

Adaptive RPython takes the generic interpreter as the input. The generic interpreter is
converted into an interpreter that includes two definitions -- one for baseline \jit{}
compilation and another is for tracing \jit{} compilation. To remove the task of manually
writing redundant definitions in
the method-traversal interpreter~\cite{Izawa:2021:ThreadedCodeGen:unpub}, we internally
generate both the method-traversal interpreter (for the baseline \jit{} compilation) and a
normal interpreter (for tracing \jit{} compilation) from the generic interpreter definition.

To convert the generic interpreter into several interpreters with two different
definitions, we need to tell adaptive \rpython{} some necessary information. For this
reason, we implement several hint functions for the generic interpreter, along with \rpython{}'s
original hints. The skeleton is shown in Listing~\ref{lst:generic_interp_def}. When developers
write the generic interpreter, they first declare an instance of \verb|JitTierDriver| class
that has a field \verb|pc|. It tells adaptive \rpython{} fundamental information such as the
variable name of the program counter. Furthermore, transforming hints should be defined in a
specific handler. \verb|can_enter_tier1_branch|, \verb|can_enter_tier1_jump|, and
\verb|can_enter_tier1_ret| tell the adaptive \rpython{}'s transformer the necessary information
to generate the method-traversal interpreter. The method-traversal interpreter requires
a particular kind of code in the \verb|JUMP_IF|, \verb|JUMP|, and \verb|RET| bytecode handlers so
that hints can be called in those handlers. The  requisite information in each handler and
hint is the following:


\paragraph{can\_enter\_tier1\_branch.} The method-traversal interpreter needs to drive the engine
to trace both sides of a branch instruction. Thus, we have to pass the following
variables/expressions to the transformer:

\begin{itemize}
\item the true and false paths of program counters: to manage them in
  the traverse stack
\item the conditional expression: to generate the \verb|if| expression.
\end{itemize}

In the handler of \verb|JUMP_IF| in Listing~\ref{lst:generic_interp_def}, we pass
the following statements/expressions: \verb|target| as \verb|true_path|,
\verb|pc+1| as a \verb|false_path| and \verb|self.is_true| as \verb|cond|. We also
remember and mark the \verb|cond| to utilize at trace-stitching inside of adaptive
\rpython{} (details are explained in
Section~\ref{sec:resoving_guard_failures_and_bridges}).

\paragraph{can\_enter\_tier1\_jump.} A jump instruction possibly placed at the end of a
function/method. In this case, the engine does not follow the destination but
takes out a program counter from the top of \verb|traverse_stack|. Otherwise, the jump
instruction performs as defined in the original interpreter . Thus, the method-traversal interpreter
requires the program counter of a jump target to manage it in the \verb|traverse_stack|.
As shown in the handler of \verb|JUMP| in Listing~\ref{lst:generic_interp_def}, we
pass \verb|target| to the \verb|transform_jump| function.

\paragraph{can\_enter\_tier1\_ret.} A return instruction is invoked at the end of a
function/method. The method-traversal interpreter requires a return value,
so we have to pass \verb|w_x| via \verb|transform_ret|, as illustrated in the handler of
\verb|RET| in Listing~\ref{lst:generic_interp_def}.

\paragraph{we\_are\_in\_tier2.} This is a hint function to tell the
area where a definition for the tracing \jit{} compilation is written to the transformer.
One thing we need to do is to specify the type of handler function defined
immediately above. For example, we add the keyword argument \verb|kind='branch'| in
the handler of \verb|JUMP_IF| in Listing~\ref{lst:generic_interp_def}.





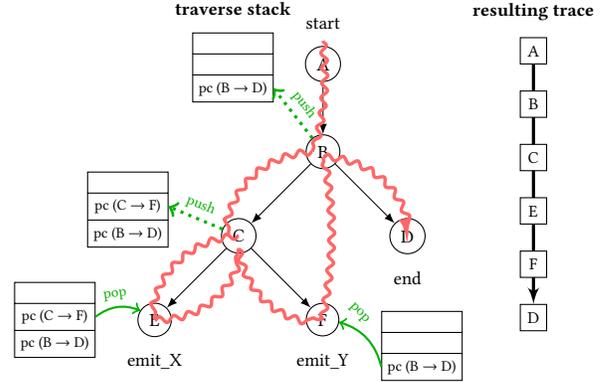
\begin{figure}[t]
  \centering
  \begin{tikzpicture}[auto,transform shape,label distance=2mm,scale=0.7,node
    distance=16mm,
    stack/.style={rectangle split, rectangle split parts=#1,draw,anchor=center,font=\footnotesize}]
    \tikzstyle{circ} = [draw,circle,fill=white]
    \tikzstyle{box} = [draw,rectangle,minimum width=5mm,minimum
    height=5mm,fill=white,align=center,font=\small]
    \tikzstyle{trace} = [red!60,line width=.45mm,decorate,
    decoration={snake,amplitude=.4mm,segment length=2mm,post length=.5mm}]

    \pgfdeclarelayer{background}
    \pgfdeclarelayer{main}
    \pgfdeclarelayer{foreground}
    \pgfsetlayers{background,main,foreground}

    \node [circ,label=above:{start}] (n1) {A};
    \node [circ,below=10mm of n1] (n2) {B};
    \node [circ,below left=of n2] (n3) {C};
    \node [circ,below left=of n3,label={below:{emit\_X}}] (n4) {E};
    \node [circ,below right=of n3,label={below:{emit\_Y}}] (n5) {F};
    \node [circ,below right=of n2,label={below:{end}}] (n6) {D};

    \draw [->,-latex] (n1) -- (n2);
    \draw [->,-latex] (n2) -- (n3);
    \draw [->,-latex] (n3) -- (n4);
    \draw [->,-latex] (n3) -- (n5);
    \draw [->,-latex] (n2) -- (n6);

    \begin{pgfonlayer}{foreground}
      \draw [trace] ($(n1.north) + (0,.1)$) to [bend left=10] ($(n2.center) - (.2,0)$);
      \draw [trace] ($(n2.center) - (.2,0)$) to [bend right=40] ($(n3.center) - (.2,0)$);
      \draw [trace] (n3.center) to [bend right=40] (n4.center);
      \draw [trace] (n4.center) to [bend right=40] ($(n3.south) + (0,.1)$);
      \draw [trace] ($(n3.south) + (0,.1)$) to [bend right=40] (n5.center);
      \draw [trace] (n5.center) to [bend right=10] ($(n2.center) - (0,0.1)$);
      \draw [-latex,trace] ($(n2.center) - (0,0.1)$) to [bend left=40] (n6.center);
    \end{pgfonlayer}

    \node [stack=3, above left=10mm of n2,label=above:{\bf traverse stack}] (stack 1) {
      \nodepart{three}{pc (B $\rightarrow$ D)}
    };

    \path [->,dotted,very thick,green!70!black] (n2)
    edge [midway,sloped] node {\footnotesize push} (stack 1.three east);

    \node [stack=3, left=10mm of n3,yshift=5mm] (stack 2) {
      \nodepart{two}{pc (C $\rightarrow$ F)}
      \nodepart{three}{pc (B $\rightarrow$ D)}
    };

    \path [->,dotted,very thick,green!70!black] (n3)
    edge [midway,sloped] node {\footnotesize push} (stack 2.two east);

    \node [stack=3, left=8mm of n4] (stack 3) {
      \nodepart{two}{pc (C $\rightarrow$ F)}
      \nodepart{three}{pc (B $\rightarrow$ D)}
    };

    \path [->,thick,green!70!black] (stack 3.two east)
    edge [midway,sloped,bend left] node {\footnotesize pop} (n4);

    \node [stack=3, right=8mm of n5,yshift=-5mm] (stack 4) {
      \nodepart{three}{pc (B $\rightarrow$ D)}
    };

    \path [->,thick,green!70!black] (stack 4.three west)
    edge [sloped,bend right] node [above left=0mm and 1mm] {\footnotesize pop} (n5);

    \node [box,right=34mm of n1,yshift=2.5mm] (a) {A};
    \node [box,below=5mm of a] (b) {B};
    \node [box,below=5mm of b] (c) {C};
    \node [box,below=5mm of c] (e) {E};
    \node [box,below=5mm of e] (f) {F};
    \node [box,below=5mm of f] (d) {D};
    \node [above=2mm of a] {\bf resulting trace};

    \begin{pgfonlayer}{background}
      \path [-latex',very thick,draw] (a) -- (d);

    \end{pgfonlayer}

  \end{tikzpicture}
  \caption{An overview of method-traversing in a program with nested branches. The
    left-hand side shows how we drive the meta-tracing engine by using the traverse stack. The
    snake line represents the trail of tracing. The right-hand side is a resulting trace.}
  \label{fig:method_traversing_and_result}
\end{figure}

\subsection{Just-in-Time Trace Stitching}
\label{sec:jit_trace_stitching_in_action}

The \emph{just-in-time trace stitching} technique builds back up the original control of a
trace yielded from the method-traversal interpreter to emit an executable \jit{}ted code. This
is the essential component in the baseline \jit{} compilation of adaptive \rpython{}.

To reconstruct the original control flow, we need to connect a correct guard operation and
bridging trace. Therefore, we should treat the destination of a ``false'' branch in a
special manner
while doing trace-stitching. In tracing \jit{} compilers, at the branching instructions,
guards are inserted.\footnote{Technically speaking, type-checking guards are inserted to
  optimize the obtained trace based on the observed run-time type information. However, we
  currently handle branching guards only and leave type optimization to the last \jit{}
  tier.} After a guard operation fails numerous times, a tracing \jit{} compiler
will trace the destination path starting from the failing guard,
that is, trace the other branch. The resulting trace from a guard failure is called a bridge,
which will be connected to the loop body. On the other hand, in trace-stitching, we
perform a sequence of generating and connecting a bridge in one go. For the sake of
ease and reducing code size, we generate a trace that essentially consists of call operations
by threaded code generation.\footnote{Note that threaded code
  generation~\cite{Izawa:2021:ThreadedCodeGen:unpub} yields
  call and guard operations. In contrast to normal tracing \jit{} compilation,
  it can reduce the code size and compilation time by 80~\% and
  60~\%.}

The left-hand side of Figure~\ref{fig:method_traversing_and_result} illustrates the
process of traversing an example target method that has nested branches. In tracing a
branch instruction, we record the other destination, for example, a program counter from B to D in
node B, in the traverse stack. For example, when tracing B and C,
we push each program counter from B to D and from C to F to the traverse stack. Then, we pop
a program counter from the traversal stack and set it to E and F. After traversing all paths,
we obtain a single trace that does not keep the structure of the target as shown in
the right-hand side of Figure~\ref{fig:method_traversing_and_result}.

\subsubsection{Resolving Guard Failures and Bridges}
\label{sec:resoving_guard_failures_and_bridges}

\begin{figure}[t]
  \centering
  \begin{tikzpicture}[auto,transform shape,scale=0.8,remember picture,label distance=3mm,
    font=\small,
    stack/.style={rectangle split, rectangle split parts=#1,draw,anchor=center,font=\footnotesize},
    ]
    \tikzstyle{dot}= [circle,fill=blue!70,inner sep=0pt, outer sep=0pt]
    \tikzstyle{box} = [rectangle,fill=blue!70,inner sep=0pt, outer sep=0pt]
    \tikzstyle{lbl} = [anchor=west,align=left,text width=20mm]
    \tikzstyle{tr} = [draw,rectangle,inner sep=2pt,minimum width=18pt,pattern=north west
    lines,pattern color=gray]
    \tikzstyle{br} = [tr,pattern=bricks]
    \tikzstyle{callout} = [rectangle callout,fill=orange!70,inner sep=4pt,align=left,text
    width=10mm,]
    \tikzstyle{callout white} = [callout,fill=white]

    \node [box,minimum width=12pt,minimum height=8pt] at (0,0) (top) {};
    \node [lbl,right=2.5mm of top] (l top) {\circled{A} start};
    \node [dot,minimum size=10pt,below=7.5mm of top,] (g1) {};
    \node [lbl,right=2.75mm of g1] (l g1) {\circled{B} guard 1};
    \node [dot,minimum size=10pt,below=7.5mm of g1] (g2) {};
    \node [lbl,right=2.75mm of g2] (l g2) {\circled{C} guard 2};
    \node [box,minimum width=12pt,minimum height=8pt,below=of g2] (e1) {};
    \node [lbl,right=2.5mm of e1] (l e1) {\circled{E} emit\_X};
    \node [box,minimum width=12pt,minimum height=8pt,below=of e1,] (e2) {};
    \node [lbl,right=2.5mm of e2] (l e2) {\circled{F} emit\_Y};
    \node [box,minimum width=12pt,minimum height=8pt,below=of e2] (bot) {};
    \node [lbl,right=2.5mm of bot] (l end) {\circled{D} end};

    \draw [->,very thick,red!90!black] (g1.south west) to [bend right=60,midway] node (f1)
    {} (e2.north west);
    \draw [->,very thick,red!90!black] (g2.south west) to [bend right=60,midway] node (f2)
    {} (e1.north west);

    \node [stack=3,left=15mm of g1,yshift=5mm,
    label={[label distance=0mm]above:{\bf guard failure stack}}] (stack1) {
      \nodepart{three}{guard failure (g1)}
    };

    \node [stack=3, left=15mm of g2,yshift=-10mm] (stack2) {
      \nodepart{one}{}
      \nodepart{two}{guard failure (g2)}
      \nodepart{three}{guard failure (g1)}
    };

    \path [->,dotted,green!70!black,very thick] ($(g1.west) - (.2,0)$)
     edge [midway,sloped] node {push} ($(stack1.three east)$);

    \path [->,dotted,green!70!black,very thick] ($(g2.west) - (.2,0)$)
    edge [midway, sloped] node {push} ($(stack2.two east)$);

    \node [stack=3, right=2.5mm of l e1,yshift=10mm] (stack3) {
      \nodepart{two}{guard failure (g2)}
      \nodepart{three}{guard failure (g1)}
    };

    \draw [->,thick,green!70!black] (stack3.two) to [bend right=45,] node {pop} (l e1);

    \node [stack=3, right=2.5mm of l e2] (stack4) {
      \nodepart{three}{guard failure (g1)}
    };

    \draw [->,thick,green!70!black] (stack4.three) to [bend left=45] node {pop} (l e2);

    \begin{pgfonlayer}{background}
      \draw [->,very thick] (top) -- (g1) -- (g2) -- (bot);
      \node [tr,fit={(top) (g1) (g2) ($(e1.north)$)}] {};
      \node [br,fit={($(e1) - (0,0.1)$) ($(e2) + (0,0.1)$)}] {};
      \node [br,fit={($(e2.south)$) (bot)}] {};
    \end{pgfonlayer}

    \matrix [below left,anchor=north west, xshift=2.5mm, yshift=8mm,
    cells={nodes={font=\footnotesize,anchor=west}},inner sep=2pt]
    at (current bounding box.south west) {
      \node [tr] {}; & \node {loop body}; \\
      \node [br] {}; & \node {bridge}; \\
    };

  \end{tikzpicture}
  \caption{An overview of just-in-time trace-stitching. This shows how we resolve the
    relations between guard failures and bridges, here taking
    Figure~\ref{fig:method_traversing_and_result} as an example.}
  \label{fig:trace_stitching_with_fdstack}
\end{figure}
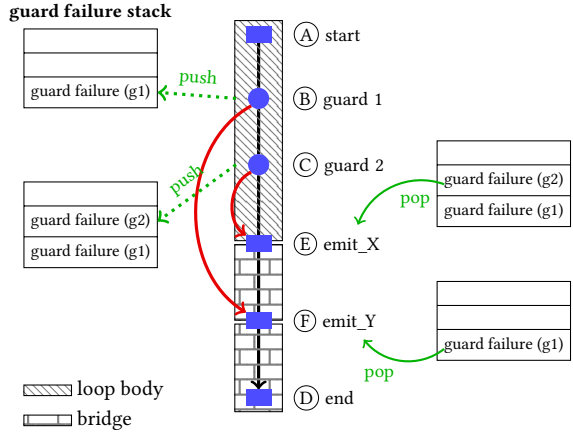

Intending to resolve the relations between guards and bridges, we utilize the nature of
the method-traversal interpreter: it manages the base-program's branches as a stack data
structure, so the connections are first-in-last-out. Thus, we implement \emph{guard failure
  stack} to manage each guard's failure. The guard failure stack saves guard failures in
each guard operation and pops them at the start of a bridge, that is, right after
\verb|emit_jump| or \verb|emit_ret| operations that indicate a cut-here line.

Before explaining the details of the algorithm, we give a high level overview of
how we resolve the connections between guard failures and bridges by using
Figure~\ref{fig:trace_stitching_with_fdstack} as an example.
First, the necessary information is which guard failure goes to which
bridges. We resolve it by sequentially reading and utilizing the guard failure stack.
When we start to stitch the trace shown in Figure~\ref{fig:trace_stitching_with_fdstack},
we sequentially read the operations from the trace. In node B, when it turns out that
the guard operation is already marked as \verb|cond| in the generic interpreter, we take its
\verb|guard failure (g1)| and push it to the guard failure stack. In node C, we do the same
thing in the case of node B. Next, in node E, we finish tracing an operation and cut the
current trace. In addition, we pop a guard failure and connect it to the bridge that we
are going to retrieve, because the bridges are lined up below with a depth-first
approach. In node F, we do the same thing in the case of node F. In node D, finally, we
finish reading and produce one trace and two bridges. The connections are illustrated as
red arrows in Figure~\ref{fig:trace_stitching_with_fdstack}.

\subsubsection{The Mechanism of Just-in-Time Trace-Stitching}
\label{sec:mechanism_trace_stitching}

How \jit{} trace-stitching works is explained through
Algorithm~\ref{alg:dotracestitching}. First, the trace-stitcher \dotracestitching{}
prepares an associative array called \verb|token_map|, and the (key, value) pair is
(\texttt{program counter}, \texttt{target\_token}).\footnote{
  \texttt{target\_token} is an identifier for a trace/bridge. When we encounter
  \texttt{emit\_jump} or \texttt{emit\_ret} operations, we get the program counter (key)
  passed as an argument to \texttt{emit\_jump} or \texttt{emit\_ret} operations, hence
  creating a new \texttt{target\_token} (value).}
Next, it declares \verb|guard_failure_stack|, \verb|trace|, and
\verb|result|. \verb|guard_failure_stack| is a key data structure to resolve the relation
between guard failures and bridges. \verb|trace| temporarily stores a handled
operation, and \verb|result| memorizes a pair of a \verb|trace| and its corresponding
\verb|guard_falure|. After done this, it manipulates each operation in the given \verb|ops|. We
specifically handle the following operations: (1) a guard operation marked by the generic
interpreter, (2) pseudo call operations \verb|CallOp(emit_ret)| and
\verb|CallOp(emit_jump)|, and (3) \rpython{}'s return and jump operations. Note that almost
all operations except guards are represented as a call operation because the resulting trace
is produced from our threaded code generator.

\paragraph{Marked Guard Operation.}

To resolve the guard--bridge relation, we firstly collect the guard operations that we have
marked in the generic interpreter. The reason we mark some guards is to distinguish between
branching guards and others. When we encounter such a guard operation,
we take its \verb|guard_failure| and append it to the \verb|guard_failure_stack|. Its
algorithm is shown in the first branching block of Algorithm~\ref{alg:dotracestitching}.

\paragraph{Pseudo call operations.}

Pseudo functions \verb|emit_jump| and \verb|emit_ret| are used as a sign of
cutting the trace at this position and to start recording a bridge. They are
represented as \verb|Call(emit_jump)| and
\verb|Call(emit_ret)| in a trace.

In the case of \texttt{CallOp(emit\_jump)}:

\begin{enumerate}
\item look up a \verb|target_token| from the \verb|token_map| using the target program
  counter as a key
\item try to retrieve a \verb|guard_failure| from the \verb|guard_failure_stack|. If it
  the stack is empty, we do nothing because the current recording trace is a body (not a
  bridge). Otherwise, pop a guard failure from the \verb|guard_failure_stack|.
\item create a jump operation with the \verb|inputargs| and \verb|target_token| and
  append to the \verb|trace|.
\item append the pair of the \verb|trace| and \verb|guard_failure| to the \verb|result|.
\end{enumerate}

In the case of \texttt{CallOp(emit\_ret)}:

\begin{enumerate}
\item take a return value (\verb|retval|) from the operation.
\item take a \verb|guard_failure| if it is not the first time.
\item create a return operation with the \verb|retval| and append to the \verb|trace|.
\item append the pair of the \verb|trace| and \verb|guard_failure|.
\end{enumerate}

\paragraph{RPython's Jump and Return.}

These operations are placed at the end of a trace. When we read them, we append the pair
of the operation and retrieve the guard failure to the \verb|result|, and then, we have
finished reading.

\section{Preliminary Evaluation}
\label{sec:preliminary_eval}

In this section, we evaluate our implementations of \jit{} trace-stitching and baseline
\jit{} compilation can actually work. Because our work is not finished, these results can
only be preliminary.

\subsection{Implementation}


We wrote a small interpreter called \tla{} in adaptive \rpython{} and executed several
micro benchmark programs on it. In the \tla{} language, we have both primitive and object
types, and they are dynamically typed.

\paragraph{Simulating Multitier Compilation.}

To conduct this preliminary evaluation, we partially implemented two-level \jit{}
compilation by separately defining interpreters for each execution tier -- baseline
\jit{} compilation (tier 1), and tracing \jit{} compilation (tier 2). In other words, to
support multitier \jit{} compilation, we prepared two interpreters that have
different jitdrivers in \tla{}; the one is for baseline \jit{} compilation while the other
is for tracing \jit{} compilation. For example, during executing a base-program, we call
separate interpreters for different \jit{} compilation tiers that are manually specified
in a base-program.

\paragraph{Accessibility}

The implementations of proof-of-concept adaptive \rpython{} and \tla{} interpreter is
hosted on Heptapod.\footnote{\url{https://foss.heptapod.net/pypy/pypy/-/tree/branch/threaded-code-generation/rpython/jit/tl/threadedcode}}
Moreover, the generic interpreter implementations are hosted on
our GitHub organization.\footnote{\url{https://github.com/prg-titech/mti_transformer}}

\subsection{Targets}

To verify that our \jit{} trace-stitching mechanism works on a program with some complex
structure, we wrote a single loop program \textbf{loop} and nested loop program
\textbf{loopabit} in \tla{} for the experiments.

Furthermore, to confirm the effectiveness of shifting different \jit{} compilation levels,
we wrote \textbf{callabit}, which has two different methods and; each method has a single
loop. According to the specified \jit{} compilation strategy, callabit has the following
variants:

\begin{enumerate}[label=(\alph*)]
\item \textbf{callabit\_baseline\_interp}: the main method is compiled by baseline \jit{}
  compilation, but the other is interpreted.
\item \textbf{callabit\_baseline\_only}: the two methods are compiled by baseline \jit{}
  compilation.
\item \textbf{callabit\_baseline\_tracing}: the main method is compiled by baseline \jit{}
  compilation, and the other is compiled by tracing \jit{} compilation.
\item \textbf{callabit\_tracing\_baseline}: the main method is compiled by tracing \jit{}
  compilation, and the other is by baseline \jit{} compilation.
\item \textbf{callabit\_tracing\_only}: all methods are compiled by tracing \jit{}
  compilation.
\end{enumerate}

All bytecode programs used for this evaluation are shown in
Appendix~\ref{sec:programs_pre_eval}.

\subsection{Methodology}

We took two kinds of data: the times of stable and startup speeds. When we measured the stable
speed, we discarded the first iteration and accumulated the elapsed time of each 100
iterations. In contrast, we measured the startup speed; we iterated 100 times the spawning of an
execution process. In addition, we took how many operations they emit and how much time they
consumed in tracing and compiling in the case of callabit programs. Note that in every
benchmark, we did not change the default threshold of the original RPython to enter \jit{}
compilation.

We conducted the preliminary evaluation in the following environment: CPU: Ryzen 9 5950X,
Mem: 32 GB DDR4-3200MHz, OS: Ubuntu 20.04.3 LTS with a 64-bit Linux kernel
5.11.0-34-generic.

\subsection{Result}

\paragraph{Compiling Single and Nested Loops}
\label{sec:res_single_nested_loops}

Figure~\ref{fig:vis_loopabit} visualizes the resulting traces from loopabit.tla by
baseline \jit{} compilation. We can initially confirm that our baseline \jit{} compilation
works correctly when looking at this figure.
Figure~\ref{fig:res_loop_loopabit_stable} and~\ref{fig:res_loop_loopabit_startup} show the
results of stable and startup times in the loop and loopabit programs, respectively. In stable
speed, on average, baseline \jit{} and tracing \jit{} are 1.7x and 3.25x faster than the
interpreter-only execution. More specifically, baseline \jit{} compilation is about 2x
slower than tracing \jit{} compilation. In startup time
(Figure~\ref{fig:res_loop_loopabit_startup}), baseline \jit{} compilation is about 1.9x
faster and tracing \jit{} compilation is about 5x faster. The loop and loopabit programs
are much suitable for tracing \jit{} compilation, so we consider that tracing \jit{}
compilation should be dominant in executing such programs. Furthermore, at applying the
baseline \jit{} compilation for such a program, we should reduce the value of threshold to
enter \jit{} compilation. Tuning the value is left for our future work.

\paragraph{Simulating Multitier JIT Compilation}
\label{sec:res_sim_multi_tier_jit}

Figure~\ref{fig:res_callabit} shows how the simulated multitier \jit{} compilation works on a
program with two different methods. The callabit has \verb|main| the method that repeatedly
calls \verb|sub_loop| that reduces the given number one by one. The call method is not
implemented by using jump but by invoking an interpreter, so the effectiveness of inlining by
tracing \jit{} compilation is limited in this case. In other words, \verb|main| is
relatively suitable for baseline \jit{} compilation and \verb|sub_loop| works well for tracing
\jit{} compilation.

In the stable and startup speeds (Figure~\ref{fig:res_callabit_stable}
and~\ref{fig:res_callabit_startup}), callabit\_baseline\_interp is about 3~\% slower than
the interpreter-only execution. This means that repeating back and forth between native
code and an interpreter execution leads to run-time overhead. Meanwhile, the combination
of baseline \jit{} and tracing \jit{} compilations (callabit\_baseline\_tracing) is as
fast as the tracing \jit{} compilation-only strategy (callabit\_tracing\_only). Additionally,
when looking at the Figure~\ref{fig:res_ops_time},
the baseline-tracing \jit{} strategy's trace size is about 40~\% smaller than the only
tracing \jit{} strategy. In contrast, the trace sizes are the same between
baseline-tracing \jit{} and tracing-baseline \jit{} strategies, but the tracing-baseline
\jit{} strategy is about 45~\% slower than baseline-tracing \jit{} strategy, and the
baseline-only strategy is about 5 \% faster than tracing-baseline \jit{} strategy.
From these results, we can deduce that there is a ceiling to using only a single
\jit{} strategy. Furthermore, to leverage different levels of \jit{} compilations, we have
to apply an appropriate compilation according to the structure or nature of the target
program.

In summary, our baseline \jit{} compilation is about 1.77x faster than the
interpreter-only execution in both the stable and startup speeds.\footnote{We calculated the
  geometric mean of loop, loopabit, and callabit\_baseline\_only in both stable and
  startup speeds.} Moreover, our baseline \jit{} compilation is only about 43 \% slower
than the tracing \jit{} compilation, even though it has very few optimizations, such as
inlining and type specialization. This means that our approach to enabling baseline \jit{}
compilation alongside with tracing \jit{} compilation has enough potential to work as a
startup compilation if we carefully adjust the threshold to enter a baseline \jit{}
compilation. This is left as future work.

\section{Related Work}
\label{sec:related_work}

Both well-developed \vm{}s, such as Java \vm{} or JavaScript \vm{}, and research-oriented
\vm{}s of a certain size support multitier \jit{} compilation to balance among the
startup speed, compilation time, and memory footprint. As far as the authors know, such \vm{}s
build at least two different compilers to realize multitier optimization. In contrast,
our approach realizes it in one engine with a language implementation framework.

The Java \hotspot{} \vm{} has the two different compilers, that is
C1~\cite{Kotzmann:2008:10.1145/1369396.1370017} and C2~\cite{paleczny2001java}, and four
optimization levels. The typical path is moving through the level 0, 3 to 4. Level 0 means
interpreting. On level 3, C1 compiler compiles a target with profiling information
gathered by the interpreter. If the C2's compilation queue is not full and the target
turns out to be hot, C2 starts to optimize the method aggressively (level 4). Level 1 and 2
are used when C2's compilation queue is full, or level 3 optimization cannot work.

The Firefox JavaScript \vm{} called SpiderMonkey~\cite{spidermonkey} has several
interpreters and compilers to enable multitier optimization. For interpreters, it has
normal and baseline interpreters~\cite{spidermonkeydoc}. The baseline interpreter supports
inline caches~\cite{Smalltalk80:10.1145/800017.800542,10.1007/BFb0057013} to improve its
performance. The baseline \jit{} compiler uses the same inline caching mechanism, but it
translates the entire bytecode into machine code. In addition, a full-fledged compiler
WarpMonkey~\cite{warpmonkey} compiles a hot spot into fast machine code. Besides such a
JavaScript engine, the SpiderMonkey \vm{} has an interpreter and compiler called
WASM-Baseline and WASM-Ion.

Google's JavaScript engine V8, which is included in the Chrome browser, also supports a
multitier compilation mechanism~\cite{ignition}. V8 sees it as a problem that the
\jit{}-compiled code can consume a large amount of memory, but it runs only once. The baseline
interpreter/compiler is called Ignition, and it is so highly optimized to collaborate with
V8's \jit{} compiler engine Turbofan. It can reduce the code size up to 50 \% by
preserving the original.

Google's V8 has another optimizing compiler called Liftoff~\cite{liftoff}. The Liftoff
compiler is designed for a startup compiler of WebAssembly and works alongside Turbofan.
Turbofan is based on its intermediate representation (\textsc{ir}), so it needs to
translate WebAssembly code into the \textsc{ir}, leading to a reduction in the startup
performance of the Chrome browser. However, Liftoff instead directly compiles WebAssembly
code into machine code. The liftoff compiler is tuned to quickly generate memory-efficient
code to reduce the memory footprint at startup time.

The Jikes Java Research VM (originally called
Jalape\~{n}o)~\cite{Alpern:10.1145/320384.320418}, which was developed by IBM
Research, is a research-oriented \vm{} that is written in Java. It has baseline and
optimizing \jit{} compilers and supports an optimization strategy in three-tires.

\section{Conclusion and Future Work}
\label{sec:conclusion_future_work}

In the current paper, we proposed the concept and initial stage implementation of adaptive
\rpython{}, which can generate a \vm{} that supports two-tier compilation.  In realizing
adaptive \rpython{}, we did not implement another compiler from scratch but drove the
existing meta-tracing \jit{} compilation engine by a specially instrumented interpreter
called the generic interpreter. The generic interpreter supports a fluent \textsc{api} that
can be easily integrated with \rpython{}'s original hint function.  The adaptive
\rpython{} compiler generates different interpreters that support a different compilation
tier. The \jit{} trace-stitching reconstructs the initial control flow of a generated
trace from a baseline \jit{} interpreter to emit the executable native code. In our
preliminary evaluation, when we manually applied a suitable compilation depending on the
control flow of a target method, we confirmed that the baseline-tracing \jit{} compilation
runs as fast as the tracing \jit{}-only compilation and reduces 50 \% of the trace size. From
this result, selecting an appropriate compilation strategy according to a target program's
control flow or nature is essential in the multitier compilation.

To implement an internal graph-to-graph conversion of the generic interpreter in
\rpython{} is something we plan to work on next. We currently implement the
generic interpreter transformer as source to source because it is a
proof-of-concept. For a smoother integration with \rpython{}, we need to switch
implementation strategies in the future.

To realize the technique to automatically shift \jit{} compilation tiers in Adaptive
\rpython{}, we also need to investigate a compilation scheme including of suitable
heuristics regarding when to go from one tier to the next.

Finally, we would implement our Adaptive \rpython{} techniques in the PyPy programming
language because it brings many benefits. For example, we can obtain a lot of data by running
our adaptive \rpython{} on existing polished benchmark programs to determine a certain
threshold to switch a \jit{} compilation. Furthermore, we could potentially bring
our research results to many Python programmers.

\begin{acks}
  We would like to thank the reviewers of the PEPM 2022 workshop for their valuable
  comments. This work was supported by JSPS KAKENHI grant number 21J10682 and JST ACT-X
  grant number JPMJAX2003.
\end{acks}

\bibliographystyle{ACM-Reference-Format}
\bibliography{main}

\appendix

\newpage

\section{The Algorithm of Just-in-Time Trace-Stitching}

\begin{algorithm}[h]
  \scriptsize
  \caption{\textsc{DoTraceStitching}(inputargs, ops)}\label{alg:dotracestitching}
  \SetKwProg{Fn}{Function}{:}{}

  \SetKwFunction{CreateTokenMap}{CreateTokenMap}
  \SetKwFunction{HandleEmitJump}{HandleEmitJump}
  \SetKwFunction{HandleEmitRet}{HandleEmitRet}

  \SetKwFunction{GetRetVal}{GetRetVal}
  \SetKwFunction{GetGfail}{GetGuardFailure}
  \SetKwFunction{PopGfail}{PopGuardFailure}
  \SetKwData{tokenmap}{token\_map}
  \SetKwData{fstack}{guard\_failure\_stack}
  \SetKwData{gfail}{guard\_failure}
  \SetKwData{trace}{trace}
  \SetKwData{result}{result}

  \SetKw{break}{break}

  \SetKwInOut{Input}{input}\SetKwInOut{Output}{output}
  \Input{red variables $inputargs$ of the given trace}
  \Input{a list of operations $ops$ taken from the given trace}

  \tcc{Note that token\_map and \fstack are global variables}
  \tokenmap $\leftarrow$ \CreateTokenMap(ops)\;
  \fstack, \trace, \result $\leftarrow$ [], [], []\;
  \For{op in ops}{
    \If{op is guard and marked}{
      \gfail $\leftarrow$ \GetGfail(op)\;
      append guard\_failure to \fstack\;
    }
    \ElseIf{op is call}{
      \If{op is CallOp(emit\_jump)}{
        $\trace, \gfail  \leftarrow \HandleEmitJump(op, inputargs)$\;
        append $(\trace, \gfail)$ to \result\;
        $\trace \leftarrow []$\;
      }
      \ElseIf{op is CallOp(emit\_ret)}{
        $\trace, \gfail \leftarrow \HandleEmitRet(op)$\;
        append $(\trace, \gfail)$ to \result\;
        $\trace \leftarrow []$\;
      }
      \Else{
        append $op$ to \trace\;
      }
    }
    \ElseIf{op is JumpOp}{
      append $op$ to \trace\;
      \gfail $\leftarrow$ \PopGfail()\;
      append (\trace, \gfail) to \result\;
      \break\;
    }
    \ElseIf{op is RetOp}{
      append $op$ to \trace\;
      \gfail $\leftarrow$ \PopGfail()\;
      append (\trace, \gfail) to \result\;
      \break\;
    }
    \Else{
      append $op$ to \trace\;
    }
  }

  \Return{\result}\;

  \SetKwFunction{PopGF}{PopGuardFailure}

  \Fn{\PopGF{}}{
    \If{first pop?}{
      \Return{None}\;
    }
    \Else{
      $failure \leftarrow$ pop the element from \fstack\;
      \Return{$failure$}\;
    }
  }

  \SetKwFunction{HandleEJ}{HandleEmitJump}
  \SetKwFunction{GetPC}{GetProgramCounter}

  \Fn{\HandleEJ{$op, inputargs$}}{
    $target$ $\leftarrow$ \GetPC(op)\;
    $token$ $\leftarrow$ \tokenmap[$target$]\;
    $\gfail$ $\leftarrow$ \PopGfail()\;
    append $JumpOp(args, token)$ to $\trace$\;

    \Return{$\trace$, $\gfail$}\;
  }

  \SetKwFunction{HandleER}{HandleEmitRet}

  \Fn{\HandleER{$op$}}{
    $retval \leftarrow \GetRetVal(op)$\;
    $\gfail \leftarrow \PopGfail()$\;
    append $RetOp(retval)$ to \trace\;
    \Return{\trace, \gfail}\;
  }

\end{algorithm}

\newpage

\section{Results of the Preliminary Evaluation}
\label{sec:results_pre_eval}

\begin{figure}[t]
  \centering
  \includegraphics[width=\linewidth]{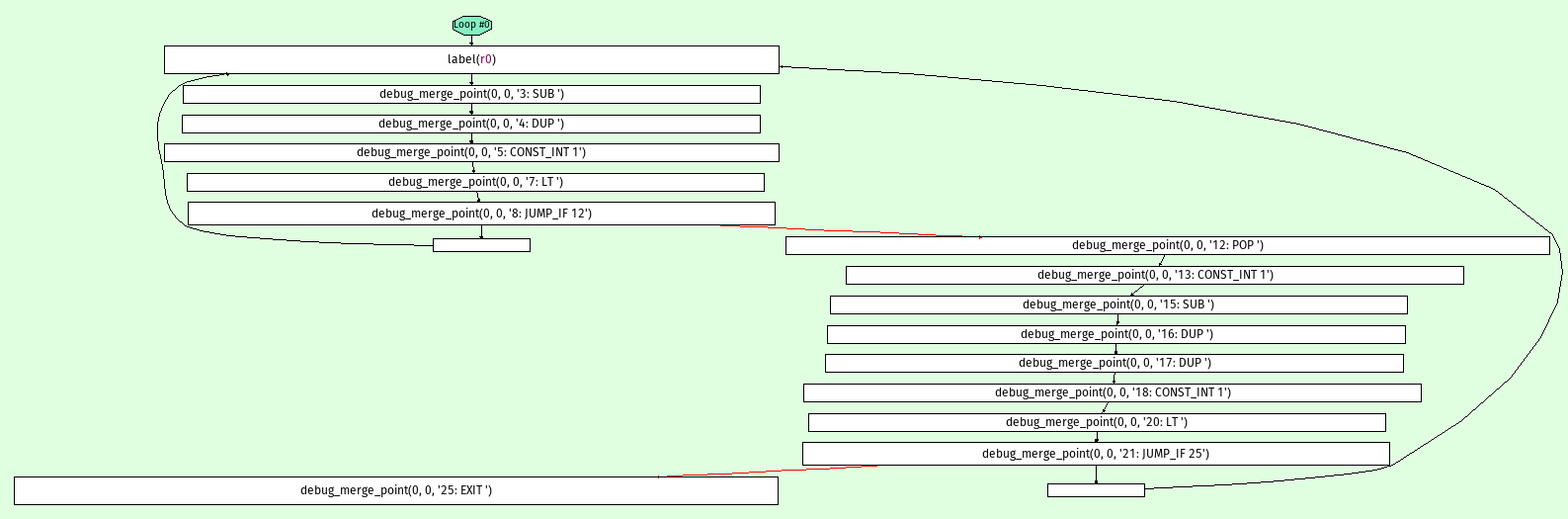}
  \caption{The visualization of the resulting traces from loopabit.tla compiled by
    baseline \jit{} compilation. Note that each trace was joined at one compile time.}
  \label{fig:vis_loopabit}
\end{figure}

\begin{figure}[h]
  \centering

  \begin{subfigure}[b]{\linewidth}
    \centering
    \includegraphics[width=.8\linewidth]{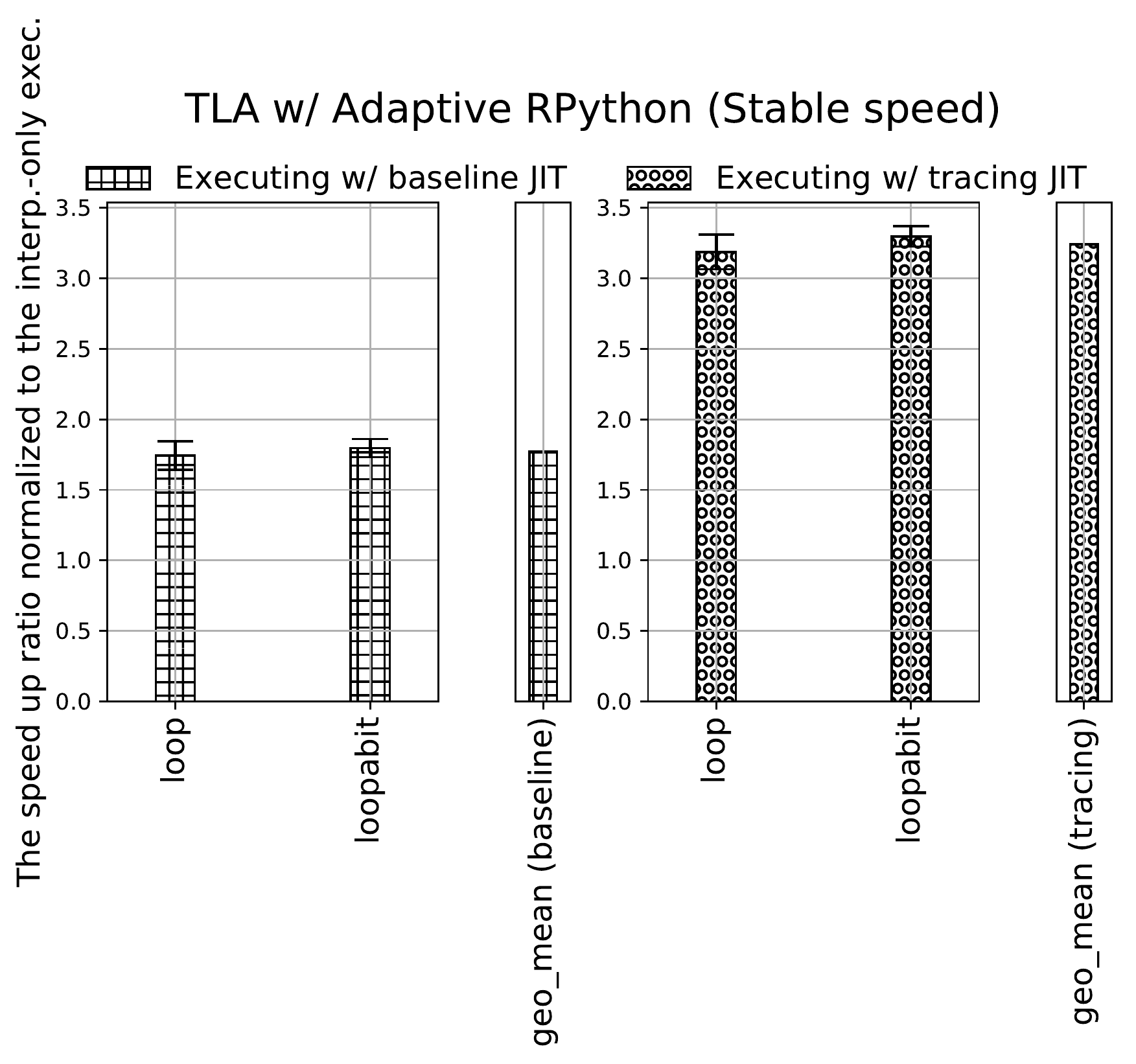}
    \caption{The result of the stable speeds of loop and loopabit.}
    \label{fig:res_loop_loopabit_stable}
  \end{subfigure}
  \begin{subfigure}[b]{\linewidth}
    \centering
    \includegraphics[width=.8\linewidth]{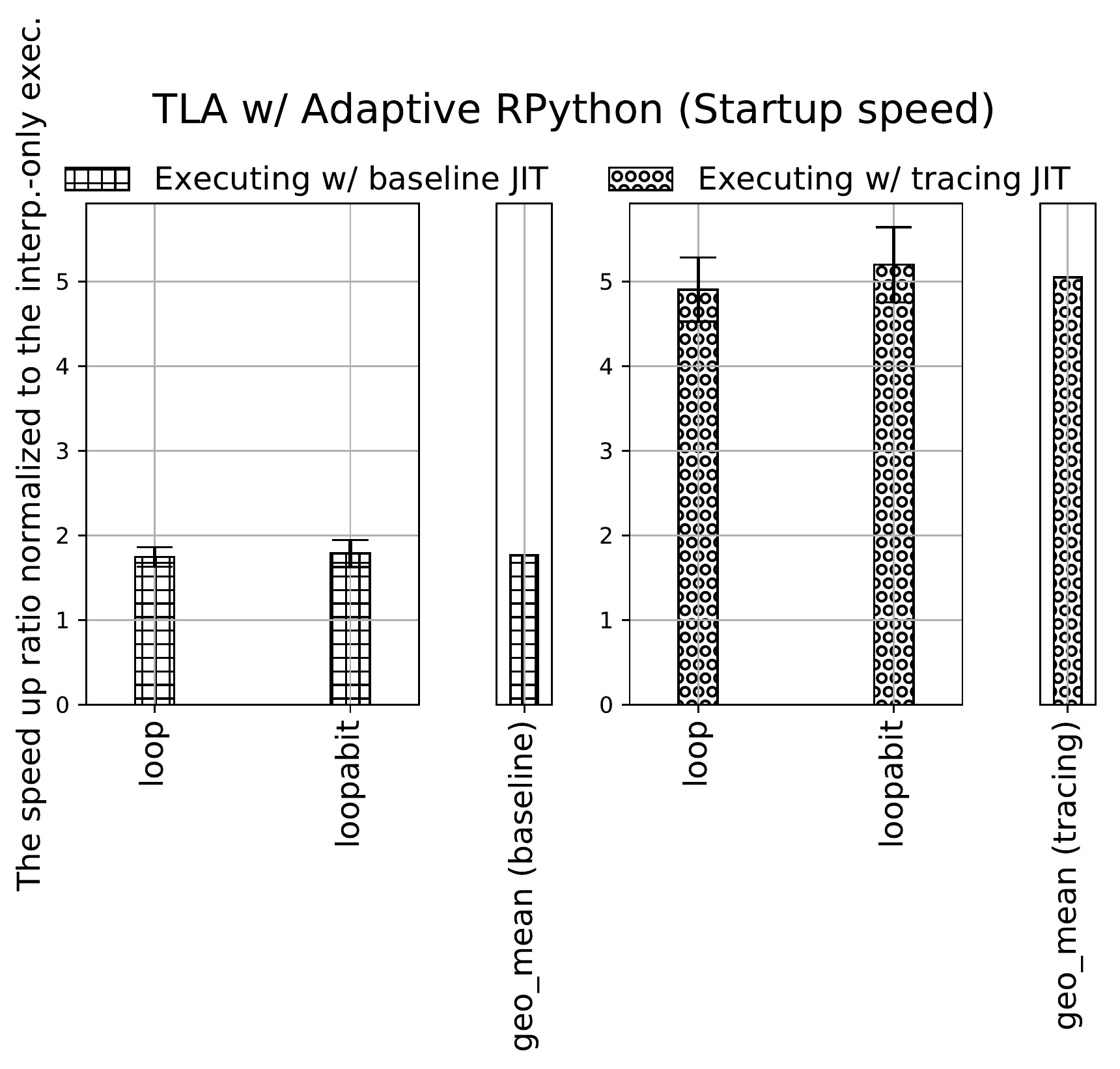}
    \caption{The result of the startup speeds of loop and loopabit. }
    \label{fig:res_loop_loopabit_startup}
  \end{subfigure}
  \caption{The results of loop and loopabit. Execution times are normalized to the
    interpreter-only execution. Higher is better}
  \label{fig:res_loop_loopabit}
\end{figure}

\begin{figure}[h]
  \begin{subfigure}[b]{\linewidth}
    \centering
    \includegraphics[width=.8\linewidth]{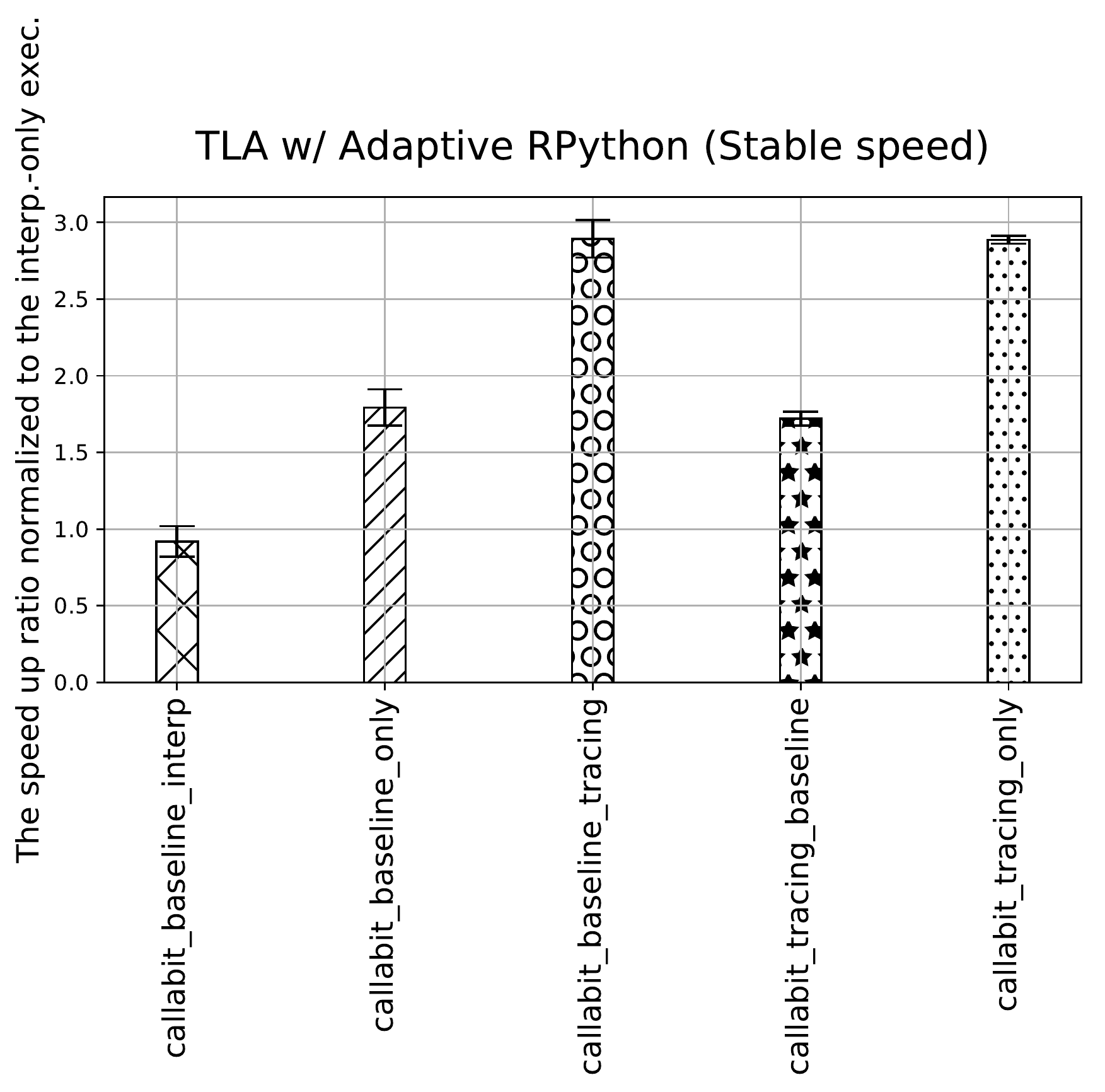}
    \caption{The result of the stable speeds of callabit programs.}
    \label{fig:res_callabit_stable}
  \end{subfigure}
  \begin{subfigure}[b]{\linewidth}
    \centering
    \includegraphics[width=.8\linewidth]{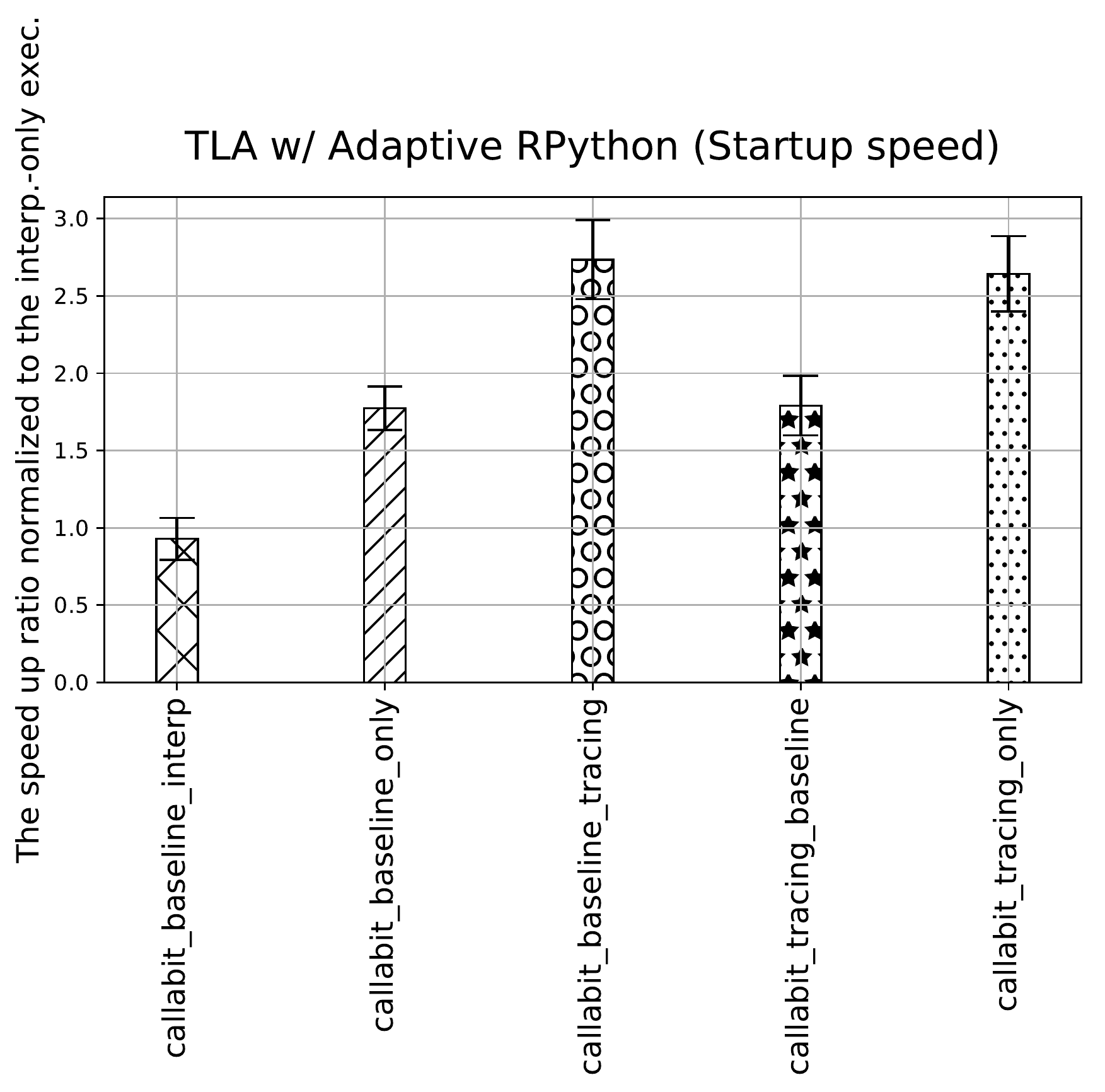}
    \caption{The result of the startup speeds of callabit programs.}
    \label{fig:res_callabit_startup}
  \end{subfigure}
  \caption{The results of callabit programs with simulated multi-tier \jit{} compilation.
    Every data is normalized to the interpreter-only execution. Higher is better.}
  \label{fig:res_callabit}
\end{figure}

\begin{figure}[h]
  \centering
  \includegraphics[width=.8\linewidth]{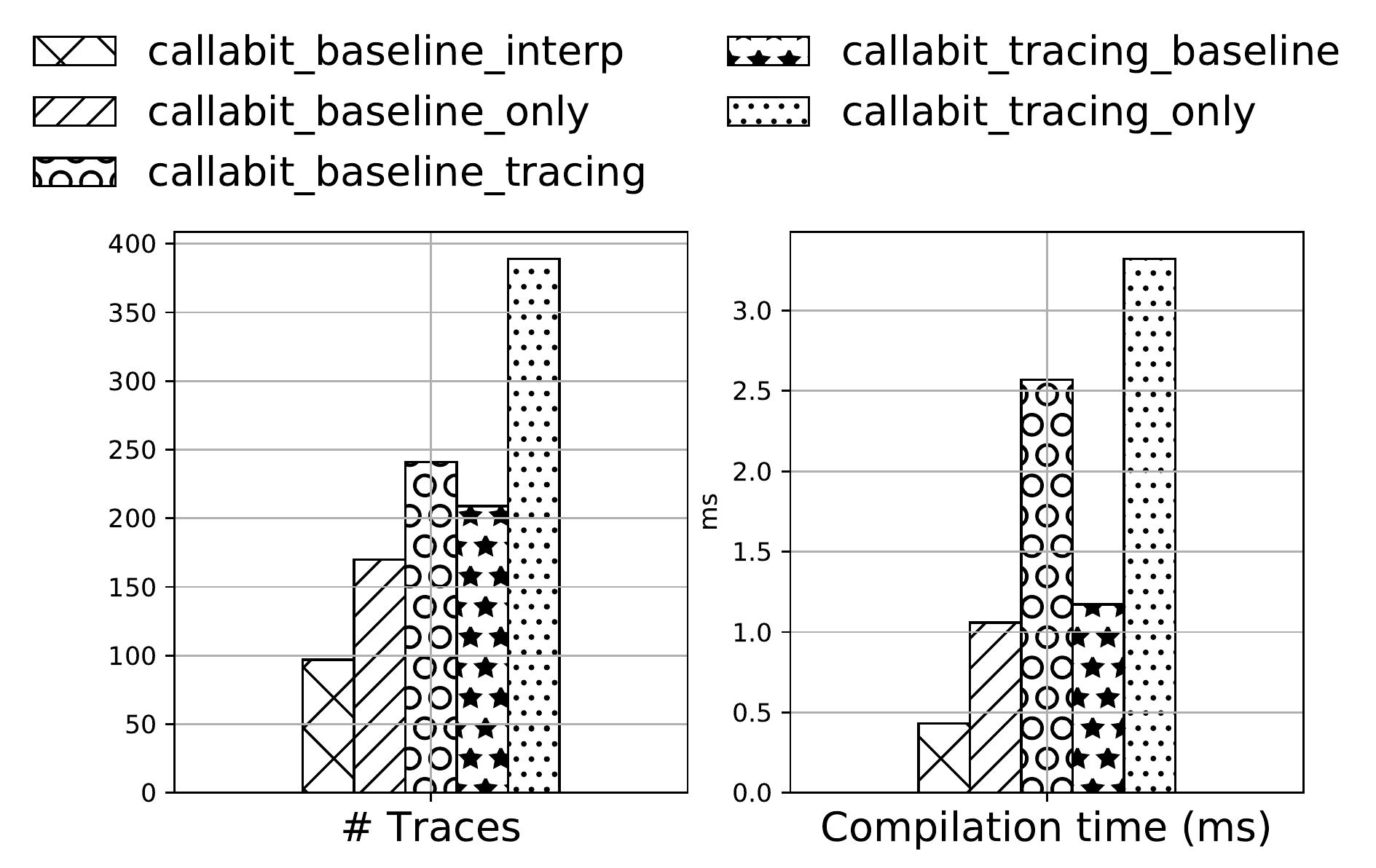}
  \caption{The trace sizes and compilation times in callabit programs. The program is
    so small that the compilation time is at most 3 \% of the total.}
  \label{fig:res_ops_time}
\end{figure}

\newpage

\section{Programs}

\subsection{The Definition of Traverse Stack}
\label{sec:def_traverse_stack}

\begin{lstlisting}[language=python, basicstyle=\tiny, caption={The definition of \texttt{traverse\_stack}.}]
class TraverseStack:
    _immutable_fields_ = ['pc', 'next']

    def __init__(self, pc, next):
        self.pc = pc
        self.next = next

    def t_pop(self):
        return self.pc, self.next

    @elidable
    def t_is_empty(self):
        return self is _T_EMPTY

_T_EMPTY = None

@elidable
def t_empty():
    return _T_EMPTY

memoization = {}

@elidable
def t_push(pc, next):
    key = pc, next
    if key in memoization:
        return memoization[key]
    result = TraverseStack(pc, next)
    memoization[key] = result
    return result
\end{lstlisting}

\subsection{Bytecode Programs Used for Preliminary Evaluation}
\label{sec:programs_pre_eval}

\begin{lstlisting}[language=python, basicstyle=\tiny, caption={Definitions of \texttt{loop}.}, captionpos=b]
# loop.tla
tla.DUP,
tla.CONST_INT, 1,
tla.LT,
tla.JUMP_IF, 11,
tla.CONST_INT, 1,
tla.SUB,
tla.JUMP, 0,
tla.CONST_INT, 10
tla.SUB,
tla.EXIT,
\end{lstlisting}

\begin{lstlisting}[language=python, basicstyle=\tiny, caption={Definition of \texttt{loopabit}}, captionpos=b]
# loopabit.tla
tla.DUP,
tla.CONST_INT, 1,
tla.SUB,
tla.DUP,
tla.CONST_INT, 1,
tla.LT,
tla.JUMP_IF, 12,
tla.JUMP, 1,
tla.POP,
tla.CONST_INT, 1,
tla.SUB,
tla.DUP,
tla.DUP,
tla.CONST_INT, 1,
tla.LT,
tla.JUMP_IF, 25,
tla.JUMP, 1,
tla.EXIT
\end{lstlisting}

\newpage

\begin{lstlisting}[language=python, basicstyle=\tiny, caption={The definition of \texttt{callabit}.}, captionpos=b]
# callabit.tla
# - callabit_baseline_interp replaces XXX with tla.CALL_NORMAL, 16
# - callabit_baseline_tracing and callabit_traing_only replace XXX
#   with tla.CALL_JIT, 16
# main(n)
tla.DUP,
tla.CALL, 16, # XXX
tla.POP,
tla.CONST_INT, 1,
tla.SUB,
tla.DUP,
tla.CONST_INT, 1,
tla.LT,
tla.JUMP_IF, 15,
tla.JUMP, 0,
tla.EXIT,
# sub_loop(n)
tla.CONST_INT, 1,
tla.SUB,
tla.DUP,
tla.CONST_INT, 1,
tla.LT,
tla.JUMP_IF, 27,
tla.JUMP, 16,
tla.RET, 1
\end{lstlisting}

\end{document}